\documentclass{article}
\usepackage[utf8]{inputenc}
\usepackage{float}
\usepackage{jheppub}
\usepackage{amsmath}
\usepackage[english]{babel}
\usepackage{amssymb}
\usepackage{mathtools}
\usepackage{color}
\definecolor{ao}{rgb}{0.0, 0.5, 0.0}

\usepackage[caption = false]{subfig}
\usepackage{graphicx}
\usepackage{longtable}
\usepackage{tikz}
\usepackage{comment}
\usepackage{capt-of}
\usepackage{amsthm}
\usepackage{hyperref}
\hypersetup{
    colorlinks=true,
    linkcolor=blue
    }

\usepackage{bm}
\usepackage[ruled,vlined]{algorithm2e} % Per gli algoritmi

\newtheorem{remark}{Remark}

\def\0{\mbox{\tiny $0$}}
\def\1{\mbox{\tiny $1$}}
\def\2{\mbox{\tiny $2$}}
\def\3{\mbox{\tiny $3$}}
\def\4{\mbox{\tiny $4$}}
\def\5{\mbox{\tiny $5$}}
\def\6{\mbox{\tiny $6$}}
\def\7{\mbox{\tiny $7$}}
\def\8{\mbox{\tiny $8$}}
\def\9{\mbox{\tiny $9$}}

\def\sn{\mathrm{sign}}

\def\circaa{\approx}

\newcommand{\SOMMA}[2]{\displaystyle\sum\limits_{#1}^{#2}}

\long\def \beq#1\eeq {\begin{equation} #1 \end{equation}}
\long\def \beaq#1\eeaq {\begin{equation}\begin{aligned} #1 \end{aligned}\end{equation}}
\long\def \bes#1\ees {\begin{equation}\begin{split} #1 \end{split} \end{equation}}
\long\def \bea#1\eea {\begin{eqnarray} #1 \end{eqnarray}}
\long\def \bse[#1]#2\ese {\begin{subequations}\label{#1}\begin{align} #2 \end{align}\end{subequations}}

\newcommand{\si}{\sigma_i}

\title{Yet another exponential Hopfield model}

\author[a,b,c]{Linda Albanese,}
\author[d,c]{Andrea Alessandrelli,}
\author[e,f]{Adriano Barra,}
\author[g,h]{Peter Sollich}

\affiliation[a]{Istituto Nazionale di Alta Matematica Francesco Severi, P.le Aldo Moro 5, 00185, Rome, Italy}
\affiliation[b]{Dipartimento di Matematica e Fisica,  Universit\`a  del Salento, Via per Arnesano, 73100, Lecce, Italy}
\affiliation[c]{Istituto Nazionale di Fisica Nucleare, Campus Ecotekne, Via Monteroni, 73100, Lecce, Italy}
\affiliation[d]{Dipartimento di Informatica, Universit\`a di Pisa, Lungarno Antonio Pacinotti, 43, 56126, Pisa, Italy}
\affiliation[e]{Dipartimento di Scienze di Base e Applicate per l'Ingegneria, Sapienza Universit\`a di Roma, Via Scarpa 16, 00161, Rome, Italy}
\affiliation[f]{Istituto Nazionale di Fisica Nucleare, Sezione di Roma1, P.le Aldo Moro 5, 00185, Rome, Italy}
\affiliation[g]{Institute for Theoretical Physics, University of G$\Ddot{o}$ttingen,
Friedrich-Hund-Platz 1, 37077 G$\Ddot{o}$ttingen, Germany}
\affiliation[h]{Department of Mathematics, King's College London, The Strand, London, UK}

\abstract{
We propose and analyze a new variation of the so-called {\em exponential Hopfield model}, a recently introduced family of associative neural networks with unprecedented storage capacity. Our construction is based on a cost function defined through exponentials of standard quadratic loss functions, which naturally favors configurations corresponding to perfect recall. Despite not being a mean-field system, the model admits a tractable mathematical analysis of its dynamics and retrieval properties that agree with those for the original exponential model introduced by Ramsauer and coworkers. By means of a signal-to-noise approach, we demonstrate that stored patterns remain stable fixed points of the zero-temperature dynamics up to an exponentially large number of patterns in the system size. We further quantify the basins of attraction of the retrieved memories, showing that while enlarging their radius reduces the overall load, the storage capacity nonetheless retains its exponential scaling. An independent derivation within the perfect recall regime confirms these results and provides an estimate of the relevant prefactors. Our findings thus complement and extend previous studies on exponential Hopfield networks, establishing that even under robustness constraints these models preserve their exceptional storage capabilities. Beyond their theoretical interest, such networks point towards principled mechanisms for massively scalable associative memory, with potential implications for both neuroscience-inspired computation and high-capacity machine learning architectures.}

\begin{document}

\maketitle

\section{Introduction}
Within the statistical mechanics community, in the first formalization of the spontaneous pattern recognition and associative memory capabilities shown by large assemblies of neurons interacting\textit{ \`a la Hebb}, huge efforts were required in the 1980s by Amit, Gutfreund and Sompolinsky (AGS) to prove that the celebrated Hopfield model (effectively the {\em harmonic oscillator} for neural networks performing these tasks \cite{Hopfield}) was able to handle a maximal number of patterns $P$ growing linearly in the number of neurons $N$ in the neural network (i.e. $P = \alpha N$ for some $\alpha \in \mathbb{R}^+$) \cite{AGS1,AGS2}.
After the AGS milestone, based on the recognition that the Hopfield model can be viewed as a pair-wise spin glass (with the $N$ McCulloch $\&$ Pitts neurons then playing the role of Ising spins),  
it was soon pointed out
-- 
by Baldi and Venkatesh \cite{Baldi} from a computer science perspective and by Elizabeth Gardner \cite{Gardner} from the point of view of statistical mechanics
--
that $p$-spin glasses equipped with Hebbian couplings are natural generalizations of the Hopfield model to many-body interactions. 
The authors of Refs.~\cite{Baldi,Gardner} discovered that the $p$-spin maximal storage capacity was far larger than for the pair-wise Hopfield model, scaling as $P = \gamma N^{p-1}$ for some $\gamma \in \mathbb{R}^+$, as a consequence of the abundance of synaptic connections in these dense networks. For the Hopfield case $p=2$, this of course recovers the linear scaling of $P$ with $N$.
%note the collapse on the linear scaling $P \propto N^1$ when $p=2$, namely in the Hopfield limit).

In the decades since then, countless techniques have been produced in the literature to deal with and formalize the emergent properties of these {\em dense neural networks} and, nowadays, we have a satisfactory picture of them, both from a statistical mechanics \cite{Diego,FraDenso,Albert,LindaRSB,LindaSuper,LindaUnsup,Lenka,DanielinoDenso,DaniDue,Malatesta} and from a computer science \cite{Krotov1,Krotov2,transformer,DenseCapacity} perspective. 
Indeed, mainly due to the efforts of Krotov and Hopfield, dense networks -- referred to by them as \textit{dense associative memories} -- have attracted renewed interest (see e.g.\ \cite{KrotovNature,KrotovNew2,KrotovNew3}). Thus, it is not entirely surprising that in recent years the generalization towards an exponential storage capacity has been achieved. If we think of these exponential models as the sum of all the dense contributions of Hebbian $p$-spin glasses, then reaching an exponential $N$-dependence for the pattern storage capacity, while impressive, has a natural interpretation. There have been two pioneering contributions regarding such exponential models, one dealing with binary neurons \cite{ExpHop1} and the other with real-valued ones \cite{ExpHop2}. Focusing on the latter, Lucibello and M\'ezard recently studied various properties of the exponential Hopfield family by relying again on tools from glassy statistical mechanics %\textit{modi operandi} 
\cite{LucibelloMezard} (and in particular a systematic comparison with Derrida's Random Energy Model \cite{Derrida}), see also \cite{ExpHop3}. 

To complement the work of Ref.~\cite{LucibelloMezard} we focus here on networks with binary neurons. In particular, we introduce and study a variation on the theme of an exponential Hopfield model equipped with such binary neurons. While not mean-field in nature, this network allows for a very simple  mathematical analysis of its storage capacity if one focuses on the {\em perfect recall regime}.  

To estimate the maximal storage capacity within a signal-to-noise approach, we choose as initial condition for the noiseless neural dynamics -- which we show is identical to the one that would be obtained 
from the model of Ref.~\cite{ExpHop2} -- 
precisely one of the patterns themselves.
%\footnote{As a sideline we point out that the {\em zero-temperature} dynamical update rule for the evolution of neural dynamics results in plain agreement with that stemming from the original model introduced in \cite{ExpHop2}.}. 
In such zero temperature dynamics, neural activities are effectively updated by steepest descent toward a local minimum of the network's cost function,  or {\em Hamiltonian} in statistical mechanics language. By adding more and more patterns to the network, we check for the  stability of the input pattern to find out that the latter is preserved as long as $P$ does not become too (exponentially) large (Sec.~\ref{Patterns as fixed points}). 
%Then, as we work with binary neurons, the neural configuration space grows as $2^N$, thus, a $O(1)$ fraction of these configurations can play as attractors, making the inspection of their related basins of attractions rather mandatory. To accomplish this task, 

To assess the size of the basins of attraction of the stored patterns,
we provide as input for the neural dynamics a corrupted version of a pattern. The result shows that broad basins for pattern storage have a cost, namely, the larger the  size of the basins of attractions, the smaller the number of patterns that can be stored. Crucially, however, even for finite radius of the basins of attraction, the exponential scaling of the storage capacity is preserved (Sec.~\ref{multiplicative noise}). Finally, by 
considering the equilibrium conditions at zero temperature and
relying upon the perfect retrieval constraint, we provide an independent proof of the exponential storage capacity, 
including an estimate of the prefactor (Sec.~\ref{NewProof}).

Overall, while our analysis remains far from being an exhaustive picture of the function of the exponential Hopfield model we introduce, it helps to shed light on the exponential storage phenomenon of modern Hebbian neural networks.

\section{Exponential Hopfield model: signal-to-noise approach}
\label{sec:exp}
Let us consider a set of $N$ Ising neurons $\bm \sigma=\{ \si \}_{i=1, \hdots, N}$ where $\si \in \{ +1, -1\}$ and a set of $P$ binary patterns $\bm \xi=\{\xi_i^\mu\}_{i=1, \hdots, N}^{\mu=1, \hdots, P}$, distributed as independent Rademacher variables, i.e.
\begin{align}
    \mathbb{P}(\xi_i^\mu)= \dfrac{1}{2} \delta_{\xi_i^\mu, -1}+ \dfrac{1}{2} \delta_{\xi_i^\mu, +1}.
\end{align}
%whose length is the same for all of them and it is $N$. 
%\newline
Before introducing  cost and loss functions, we define control and order parameters that help us to quantify the information processing capabilities shown by this model.

%\par\medskip

The control parameters are the  noise $T=1/\beta$ in the network (which we always take as $T=0$) and the storage $\gamma$ of the network\footnote{Note that the parameter $\gamma$ fine tunes the storage within the maximum storage scaling, which will be exponential in $N$ in this context, {\em vide infra}, e.g. \eqref{eq:carico1}.} while the order parameters are the $P$ Mattis magnetizations $m_{\mu}, \ \mu \in \{1,...,P\}$, defined as
\begin{equation}
m_{\mu} := \frac{1}{N}\sum_{i=1}^N \xi_i^{\mu} \sigma_i.
\end{equation} 

%\bigskip

The cost function (or \textit{Hamiltonian} in physics terminology) $\mathcal{H}_N(\bm \sigma \vert \bm \xi)$ of the {\em Exponential Hopfield Model} that we study reads as  
\begin{align}
    \mathcal{H}_N(\bm \sigma \vert \bm \xi) :=- N\SOMMA{\mu=1}{P} e^{-N\mathcal{L}^{\mu}_N(\bm \sigma \vert \bm \xi)}  = - N\SOMMA{\mu=1}{P} e^{-\frac{1}{2}||\xi^{\mu}-\sigma||^2}  =  - N\SOMMA{\mu=1}{P} e^{N\left(m_{\mu}-1\right)}\ .
    \label{eq:hamilt_exp}
\end{align}
Here we have also introduced the $P$ standard loss function $\mathcal{L}^{\mu}_N(\bm \sigma \vert \bm \xi):=\frac{1}{2N}||\boldsymbol{\xi}^{\mu}-\boldsymbol{\sigma}||^2$ for $\mu \in \{1,...,P\}$,  to highlight that when searching for the minima of the cost function we are simultaneously extremizing the loss function (note that the latter appears in the exponent, which is extensive in $N$).

While at a first glance, the above Hamiltonian does not resemble those of the standard exponential Hebbian networks (see e.g.~\cite{ExpHop1,ExpHop3}),  once the dynamical update rules for neural activities will be derived later on, we will show  that this model is actually rather similar to the original one introduced in \cite{ExpHop2}.
Alternatively, one can Taylor expand the cost function \eqref{eq:hamilt_exp} around the {\em perfect recall} regime $m_{\mu} = 1$ and recover precisely the cost function of Krotov $\&$ Hopfield's dense associative memory~\cite{Krotov1,KrotovNature}.

\subsection{Patterns as fixed points}\label{Patterns as fixed points}

By relying upon signal-to-noise techniques, our primary goal here is to estimate the storage capacity of this network and its tolerance against (quenched) noise. To obtain this information, the first question we address is the stability of a neural configuration. We start the neural dynamics by providing as initial condition exactly one of the patterns, and study how its stability changes as a function of the number of stored patterns $P$. Then we study the robustness against perturbations by blurring the initial condition (i.e.\ the input provided to the network) for its noiseless dynamics, as this provides an indication about the amplitude of the basins of attractions and, thus, the fault tolerance of the network. 

A natural starting point, in signal-to-noise approaches, is to evaluate  the effective field acting on the $i$-th neuron. To do this we exploit the definition of the Mattis magnetization $m_\mu$ and split off the contribution of the $i$-th, namely
\begin{equation}
    m_\mu = \frac1N \sum_{j=1}^N \xi_j^{\mu}\sigma_j \Rightarrow N m_\mu= \xi_i^\mu\sigma_i+ \sum\limits_{j\neq i}^N \xi_j^\mu\sigma_j.
\end{equation}
We can then recast the cost function \eqref{eq:hamilt_exp} as
\begin{equation}
    \begin{array}{lll}
         \mathcal{H}_N(\bm\sigma \vert \bm \xi)&=&-Ne^{-N}\SOMMA{\mu=1}{P} e^{\sum\limits_{j\neq i}\xi_j^\mu\sigma_j+\xi_i^\mu\sigma_i}=-Ne^{-N}\SOMMA{\mu=1}{P} e^{\sum\limits_{j\neq i}\xi_j^\mu\sigma_j}e^{\xi_i^\mu\sigma_i}.
    \end{array}
\end{equation}
Since $\sigma_i \in \{-1, +1\}$, we have the relation 
\begin{equation}
    e^{\xi_i^\mu\sigma_i}= e^{\xi_i^\mu}\left(\dfrac{\sigma_i+1}{2}\right)-e^{-\xi_i^\mu}\left(\dfrac{\sigma_i-1}{2}\right)
\end{equation}
and can write the cost function
\begin{equation}
    \begin{array}{lll}
         \mathcal{H}_N(\bm\sigma \vert \bm \xi)&=&- N e^{-N}\SOMMA{\mu=1}{P} e^{\sum\limits_{j\neq i}\xi_j^\mu\sigma_j}\left[e^{\xi_i^\mu}\left(\dfrac{\sigma_i+1}{2}\right)-e^{-\xi_i^\mu}\left(\dfrac{\sigma_i-1}{2}\right)\right]
         \\\\
         &=&- N e^{-N}\SOMMA{\mu=1}{P} e^{\sum\limits_{j\neq i}\xi_j^\mu\sigma_j}\sinh(\xi_i^\mu)\sigma_i-N e^{-N}\SOMMA{\mu=1}{P} e^{\sum\limits_{j\neq i}\xi_j^\mu\sigma_j}\cosh(\xi_i^\mu),
    \end{array}
\end{equation}
in terms of hyperbolic sine and cosine functions. 
The prefactor of $-\sigma_i$ in the first term is naturally identified as the post-synaptic field $h_i$ acting on neuron $\sigma_i$. We therefore define  
%if we set $h_i(\bm \sigma_{\backslash i}  \vert \bm \xi)$ and $\mathcal{C}_i(\bm\sigma_{\backslash i} \vert \bm \xi)$, 
for $i=1, \hdots, N$
%Now, using the standard labels $h_i$ for the post-synaptic field acting on the neuron $\sigma_i$, if we set $h_i(\bm \sigma_{\backslash i}  \vert \bm \xi)$ and $\mathcal{C}_i(\bm\sigma_{\backslash i} \vert \bm \xi)$, for $i=1, \hdots, N$, as
\begin{equation}
    \begin{array}{lll}
         h_i(\bm\sigma_{\backslash i}|\bm \xi):= e^{-N}\SOMMA{\mu=1}{P} e^{\sum\limits_{j\neq i}\xi_j^\mu\sigma_j}\sinh(\xi_i^\mu),
         \\\\
         \mathcal{C}_i(\bm\sigma_{\backslash i} \vert \bm \xi) := e^{-N}\SOMMA{\mu=1}{P} e^{\sum\limits_{j\neq i}\xi_j^\mu\sigma_j}\cosh(\xi_i^\mu)\,,
    \end{array}
\end{equation}
where $\bm\sigma_{\backslash i}$ indicates the neural configuration $\bm \sigma$ without $\sigma_i$. In terms of these quantities we have simply
\begin{equation}
    \begin{array}{lll}
         \mathcal{H}_N(\bm\sigma \vert \bm \xi)
         &=&-N\left[\,h_i(\bm\sigma_{\backslash i}|\bm \xi)\sigma_i+\,\mathcal{C}_i(\bm\sigma_{\backslash i}|\bm\xi)\right].
         \label{eq:newH}
    \end{array}
\end{equation}
%where with $\bm\sigma_{/i}$ we meant that $\mathcal{C}(\bm\sigma_{/i}|\bm\xi)$ does not depend on the variable $\sigma_i$.
In this way we have rewritten the cost function in terms of the internal fields, or more specifically as a sum of a $\sigma_i$-independent term and one that is linear in $\sigma_i$.

From~\eqref{eq:newH} we can easily compute the cost of the spin flip $\sigma_i \to -\sigma_i$, i.e.\ the difference between \eqref{eq:newH} evaluated at $-\sigma_i$  and at $\sigma_i$. We call this quantity $\Delta E$ as in physical terms it is an energy change and have explicitly
%nalogy with the {\em energy} in Physics) and that reads as 
%In this way we have rewritten the cost function in terms of its internal fields and, in particular, as a sum of a $\sigma_i$-independent term and of a linear term in $\sigma_i$.
%Now, exploiting \eqref{eq:newH}, we can compute the cost of the spin flip $\sigma_i \to -\sigma_i$ (namely the difference between \eqref{eq:newH} evaluated at $\sigma_i$  and at $-\sigma_i$) that we call $\Delta E$ (in analogy with the {\em energy} in Physics) and that reads as 
\begin{equation}
    \begin{array}{lll}
         \Delta E=\mathcal{H}_N(\bm\sigma_{\backslash i},-\sigma_i|\bm\xi)-\mathcal{H}_N(\bm\sigma_{\backslash i},\sigma_i|\bm\xi)&=&2Nh_i(\bm\sigma_{\backslash i}|\bm \xi)\sigma_i
         \\\\
         &=&N e^{-N}\SOMMA{\mu=1}{P} e^{\sum\limits_{j\neq i}\xi_j^\mu\sigma_j}\sinh(\xi_i^\mu)\sigma_i.
    \end{array}
\end{equation}
The zero temperature neural dynamics is such that it always reduces the energy, so that if $\Delta E>0$ we have $\sigma_i^{(n+1)}=\sigma_i^{(n)}$; while if $\Delta E<0$ we have $\sigma_i^{(n+1)}=-\sigma_i^{(n)}$, giving overall
%{\em En route} toward energy minimization, if $\Delta E>0$, we have $\sigma_i^{(n+1)}=\sigma_i^{(n)}$, instead if $\Delta E<0$, we have $\sigma_i^{(n+1)}=-\sigma_i^{(n)}$, thus
\begin{equation}
    \sigma_i^{(n+1)}=\sigma_i^{(n)}\sn\left[h_i(\bm\sigma_{\backslash i}^{(n)}|\bm\xi)\sigma_i^{(n)}\right] =\sigma_i^{(n)}\sn\left[e^{-N}\SOMMA{\mu=1}{P} e^{\sum\limits_{j\neq i}\xi_j^\mu\sigma_j^{(n)}}\sinh(\xi_i^\mu)\sigma_i^{(n)}\right].
    \label{eq:s2_1}
\end{equation}
According to \eqref{eq:s2_1}, the pattern $\bm\sigma=\bm\xi^1$ is stable with regard to a flip of spin $i$ when
%Through \eqref{eq:s2_1}, we check the stability of $\bm\sigma=\bm\xi^1$ by inspecting whenever
\begin{equation}
    \Delta E=2Nh_i(\bm \xi^1_{\backslash i} \vert \bm \xi)\xi_i^1 > 0 \quad\Longrightarrow \quad e^{-N}\SOMMA{\mu=1}{P} e^{\sum\limits_{j\neq i}\xi_j^\mu\xi^1_j}\sinh(\xi_i^\mu)\xi^1_i>0.
     \label{ineq}
\end{equation}
We now exploit the Central Limit Theorem (in the limit of large $N$ and $P$) to approximate the distribution of the quantity $h_i(\bm \xi^1_{\backslash i} \vert \bm \xi)\xi_i^1$ appearing in inequality~\eqref{ineq} as Gaussian,  $\mathcal{N}(\mu_1,\sqrt{\mu_2-\mu_1^2}\;)$, with moments given by
%We exploit the Central Limit Theorem (CLT) (in the large $N,P$ limits) to approximate the argument of the sign in Eq.\eqref{eq:s2_1}  as a Gaussian  $\mathcal{N}(\mu_1,\sqrt{\mu_2-\mu_1^2}\;)$, setting
\begin{equation}
\begin{array}{lll}
     \mu_1 = \mathbb{E}\left[e^{-N}\SOMMA{\mu=1}{P} e^{\sum\limits_{j\neq i}\xi_j^\mu\xi^1_j}\sinh(\xi_i^\mu)\xi^1_i\right]=  e^{-1}\sinh(1), 
     \\\\
     \mu_2 =\mathbb{E}\left[\left(e^{-N}\SOMMA{\mu=1}{P} e^{\sum\limits_{j\neq i}\xi_j^\mu\xi^1_j}\sinh(\xi_i^\mu)\xi^1_i\right)^2\right]= e^{-2}\sinh^2(1) +  e^{-2}\sinh^2(1) P \left(\dfrac{1+e^{-4}}{2}\right)^{N-1}
\end{array}
\label{eq:momenta_exp}
\end{equation}
where the expectations are over the pattern distribution.
We can then write 
%Thus, we can write
\begin{equation}
     h_i(\bm\xi^1 \vert \bm \xi)=\mu_1+ z\sqrt{\mu_2-\mu_1^2}, \qquad z\sim\mathcal{N}(0,1)
\end{equation}
so that $h_i(\bm\xi^1 \vert \bm \xi)>0$ is equivalent to
\begin{equation}
    z >-\dfrac{1}{\sqrt{P \left(\dfrac{1+e^{-4}}{2}\right)^{N-1}}}.
\end{equation}
This happens with probability
\begin{equation}\label{xAppendix}
    \mathbb P\left(z >-\dfrac{1}{\sqrt{P \left(\dfrac{1+e^{-4}}{2}\right)^{N-1}}}\right)=1-\dfrac{1}{2} \mathrm{erfc}\left[\dfrac{1}{ \sqrt{2P \left(\dfrac{1+e^{-4}}{2}\right)^{N-1}}}\right]
\end{equation}
%which goes to $1$ as $N\to\infty$ if and only if
which goes to $1$ as $N\to\infty$ if and only if
\begin{equation}
    P \left(\dfrac{1+e^{-4}}{2}\right)^{N-1}\to 0
    \label{P_cond}
\end{equation}
 If we more specifically ask that the second term in (\ref{xAppendix}) should vanish as $N^{-a}$ for $N\to\infty$, for $a>0$, we find quantitatively the leading order bound on $P$:
%This is satisfied if, for example,
\begin{equation}
P \leq \dfrac{1}{2a\ln(N)}\left(\dfrac{2}{1+e^{-4}}\right)^{N-1}.
    \label{eq:carico1}
\end{equation}
%\begin{equation}
%    P \ll\left(\dfrac{2}{1+e^{-4}}\right)^{N-1}\Longrightarrow P {\leq} \dfrac{1}{2\log(N)}\left(\dfrac{2}{1+e^{-4}}\right)^{N-1}.
%    \label{eq:carico1}
%\end{equation}

\begin{algorithm}[tb]
\label{alg:MC_simul}
\caption{MCMC noiseless parallel Glauber dynamics \label{algo:2}}
\KwIn{Couplings $$  \{\xi^{\mu}_i\}^{\mu=1,\cdots, P}_{i=1, \cdots, N}$$ 
Input $\bm\sigma^{(0)}$, number of dynamic steps $ N_p $.}
\KwOut{Final neural configuration $\bm\sigma^{*} $}

Set $n = 0$\;

\Repeat{$n = N_p$}{
    Update the neural configuration $\bm\sigma$ according to \\$\sigma_{i}^{(n)} = \text{sign} \left[h_i\left(\bm\sigma^{(n-1)}_{\backslash i}\Big|\bm\xi\right)\right] \;\;\mathrm{for}\;\;i = 1,\cdots,N$\;
    $ n = n + 1 $\;}
\end{algorithm}

\begin{figure}
    \centering
    \includegraphics[width=12cm]{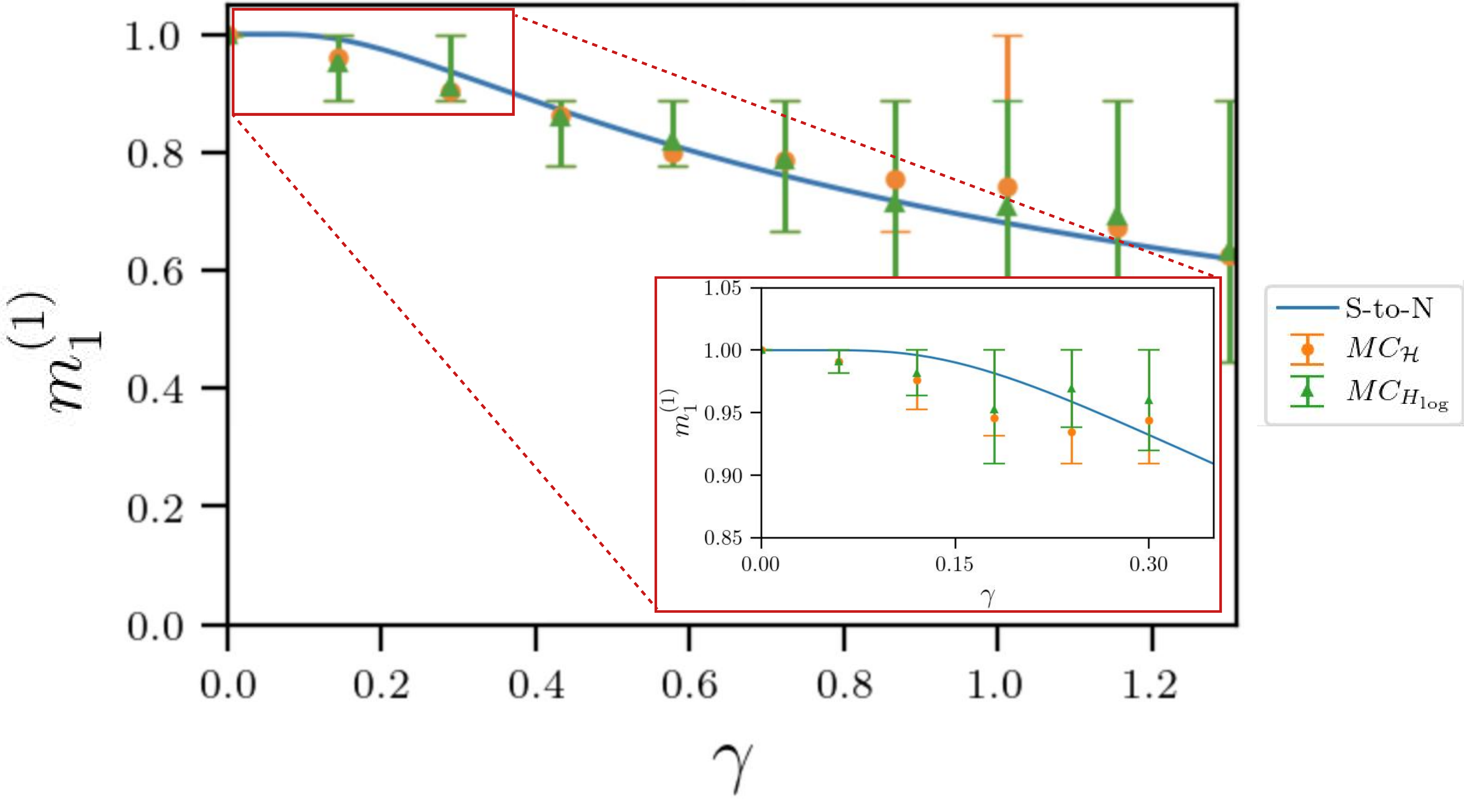}
    \caption{Mattis magnetization after one update, $m_1^{(1)}$, vs the storage $\gamma$. The blue curve shows predictions of the Mattis magnetization $m_1^{(1)}$ as a function of the storage capacity $\gamma=P\left(\frac{1+e^{-4}}{2}\right)^{N-1}$ as predicted by the signal-to-noise technique \eqref{eq:m1_2}. Orange points are MCMC outcomes (with relative error-bars) from simulations with our exponential Hopfield model \eqref{eq:hamilt_exp}, while green points (with relative error-bars) are obtained from MCMC on the original cost function \eqref{eq:H_iAyN2} from \cite{ExpHop1}. Note that, due to computational constraints, the network studied here consists of only $N=22$ neurons, yet it almost perfectly handles ($m_1^{(1)} \approx 1$) $P\approx 1.4 \cdot 10^{5}$ patterns (corresponding to the first point for nonzero $\gamma$ in the inset).
    %second point from the left in the inset plot), that is up to $\gamma=0.1$.
    }
    %\\
    %\red{@@ not entirely clear what "second" means here as the main plot shows a point at $\gamma\approx 0$ but the inset does not. I'd give the approximate value of $\gamma$ [$\gamma\approx 0.15$?] instead @@}\\
    %\blue{The possible mistake here is that the insert has a vary small point also at $\gamma \sim 0.0$ (it has  almost vanishing standard deviation thus it is difficult to be seen).}}
    \label{fig:s2n_MC_load}
\end{figure}

By the simple arguments above\footnote{The criterion for stability we have used is based on the probability that a {\em given} spin $i$ will not flip in zero temperature dynamics. Arguably one should use the stricter criterion that {\em none} of the spins $i=1,\ldots N$ will flip. We show in the Appendix that this leaves the result for the storage capacity unchanged except for prefactors. 
%It can be pointed out that this argument is somehow raw as it inspects solely the stability with respect to a single spin flip (that is reasonable only  in the {\em perfect recall} regime, where all the spin are locked together). In the Appendix we extend this analysis by studying the probability of $O(N)$ flips to show that the above picture remains however, substantially, preserved.
}, we thus already have an indication that our network is able to handle an exponential (in the number of neurons $N$) number of patterns $P$. While such an exponential storage capacity may appear surprising, it becomes more intuitive (as discussed in the introduction) once the exponential dependence on the $m_\mu$ in the Exponential Hopfield Model is expanded into an infinite series of $p$-spin interactions.
%Despite this result may appear surprising at a first glance, by another perspective this extra-storage capability is somehow expected: indeed, since the early days, both in Computer Science \cite{Baldi} and  in Statistical Mechanics \cite{Gardner}, it was clear that dense Hebbian networks\footnote{Dense Hebbian networks are  many-body $p$-spin Hopfield models in the Mattis magnetizations \cite{Burioni,Gardner2}, namely they are the natural polynomial generalization of the Hopfield reference, the latter being the simplest monomial of order $p=2$ accounting for pair-wise interactions only among neurons.} of order $p$ have  a storage capacity that scales as $P \sim N^{p-1}$. If we think that these exponential Hopfield models can be represented by expanding their cost functions \textit{\'a la Taylor} in all the dense networks (see e.g. \cite{Diego}), it is rather natural to claim that their capacity could actually become exponential. 
%Further, Monte Carlo simulations, performed following the pseudocode reported in Algorithm \ref{alg:MC_simul}, corroborate this statement as shown in Fig.~\ref{fig:s2n_MC_load}.

\begin{remark}
    If we change the scaling factor of the exponent in the cost function \eqref{eq:hamilt_exp}, i.e.,
    \begin{equation}
        \mathcal{H}_N(\bm\sigma|\bm\xi)=-N\SOMMA{\mu=1}{P}e^{N\mathcal{C}(m_\mu-1)}\;\;\;\;\mathrm{with}\;\;\mathcal{C}> 0,
    \end{equation}
    the storage capacity of the network changes accordingly, to
    \begin{equation}
        P \sim \left(\dfrac{2}{1+e^{-4\mathcal{C}}}\right)^{N-1}.
    \end{equation}
For $\mathcal{C}\to\infty$ this approaches $P\sim 2^{N-1}$. 
%We stress that if $\mathcal{C}\to\infty$ then $P\sim 2^{N-1}$.
Note, however, that for any finite $\mathcal{C}$ the number of patterns that can be stored remains an exponentially (in $N$) small fraction of the number of network configurations $2^N$, which is intuitively reasonable given that each pattern should be surrounded by a basin of attraction separating it from other patterns.
\end{remark}

Exploiting the updating rule \eqref{eq:s2_1}, it is possible to investigate how  the value of the Mattis magnetization evolves after one step , i.e.\ from $n$ to $n+1$, of the network dynamics when all spins are updated in parallel \cite{1MCMC1,1MCMC2}:
\begin{equation}
    m_1^{(n+1)}= \dfrac{1}{N}\SOMMA{i=1}{N}\xi_i^1\sigma_i^{(n+1)}=\dfrac{1}{N}\SOMMA{i=1}{N}\xi_i^1\sigma_i^{(n)}\sn\left[h_i(\bm\sigma_{\backslash i}^{(n)}|\bm\xi)\sigma_i^{(n)}\right].
\end{equation}
Now if we assume as above that initially ($n=0$) the neural configuration is perfectly aligned with $\bm\xi^1$, i.e.\ $\bm\sigma^{(0)}\equiv \bm\xi^1$, we have for the  Mattis magnetization after one step of (parallel) neural updating
\begin{equation}
    m_1^{(1)}= \dfrac{1}{N}\SOMMA{i=1}{N}\sn\left[h_i(\bm\xi^1_{\backslash i}|\bm\xi)\xi_i^1\right].
    \label{eq:up_mattis}
\end{equation}
Approximating the argument of the sign function as a Gaussian random variable as above then gives
% Using the CLT on the sums over $\mu$ and $j$, we  can replace each term $ h_i({\bm\xi}^1_{\backslash i}|{\bm\xi})$ appearing in the argument of the sign function in \eqref{eq:up_mattis},  as $\mu_1+z_i\sqrt{\mu_2}$ with $z_i \sim \mathcal{N}(0,1)$, where
% \begin{eqnarray}
% \mu_1&\coloneqq \mathbb{E}_{\xi}\left[ h_i(\bm\xi^1_{\backslash i}|\bm\xi)\xi_i^1\right],\label{eq:mu1_matt}
% \\
% \mu_2&\coloneqq \mathbb{E}_{\xi}\left\{\left[ h_i(\bm\xi^1_{\backslash i}|\bm\xi)\xi_i^1\right]^2\right\}.\label{eq:mu2_matt}
% \end{eqnarray}
% Thus, \eqref{eq:up_mattis} becomes
\begin{equation} \label{eq:bibo_matt}
    m_1^{(1)}=\dfrac{1}{N}\SOMMA{i=1}{N}\mathrm{sign}\left(\mu_1 +z_i\sqrt{\mu_2-\mu_1^2}\right)\,,
\end{equation}
% where we omit the superscript $(n+1)$ in the l.h.s. term in order to lighten the notation.\\
For large values of $N$, the arithmetic mean appearing in the r.h.s. of \eqref{eq:bibo_matt} can be replaced with the expected value by exploiting that, for a generic function $g(z)=\mathrm{sign}(\mu_1 + z \sqrt{\mu_2})$,
\begin{equation}
    \dfrac{1}{N}\SOMMA{i=1}{N} g(z_i) \;\;\mathrm{with}\;\; z_i\sim \mathcal{N}(0,1) \xrightarrow[N\to\infty]{} \mathbb{E}[g(z)]=\displaystyle\int \dfrac{dz}{\sqrt{2\pi}}e^{-\frac{z^2}{2}} g(z)\,.
    \label{eq:largemean_matt}
\end{equation}
In the large $N$ limit, we can therefore rewrite \eqref{eq:bibo_matt} as
\begin{equation}
    m_1^{(1)}=\displaystyle\int\dfrac{dz\,e^{-\frac{z^2}{2}}}{\sqrt{2\pi}} \mathrm{sign}\left(\mu_1+z\sqrt{\mu_2-\mu_1^2}\right)= \mathrm{erf}\left(\dfrac{\mu_1}{\sqrt{2(\mu_2-\mu_1^2)}}\right) \,.\label{eq:m1_1}
\end{equation}
Inserting the values of $\mu_1$ and $\mu_2$ from \eqref{eq:momenta_exp}, we finally get the prediction
%By replacing them in \eqref{eq:m1_1}, we get
\begin{equation}
    m_1^{(1)} = \mathrm{erf}\left[\left(2 P \left(\dfrac{1+e^{-4}}{2}\right)^{N-1}\right)^{-1/2}\right] \,.\label{eq:m1_2}
\end{equation}
For the pattern to be stable we require $m_1^{(1)}=1$, so we again need the argument of the error function to diverge. As expected we thus retrieve the condition \eqref{P_cond} for the storage capacity that we had obtained by considered stability against single spin flips. Eq.~\eqref{eq:m1_2} has the additional benefit that it can also be checked directly against numerical Markov Chain Monte Carlo (MCMC) simulations, performed following the pseudocode reported in Algorithm \ref{alg:MC_simul}. The results in 
Fig.~\ref{fig:s2n_MC_load} corroborate our prediction~\eqref{eq:m1_2} for the Mattis magnetization after one parallel update step, and hence our more general result of an exponential storage capacity. 

\begin{remark}
We  highlight that the  cost function
    \begin{equation}
        \tilde{\mathcal{H}}_N(\bm\sigma|\bm\xi)  = - \ln\left[\SOMMA{\mu=1}{N}e^{N\,m_\mu}\right],
        \label{eq:H_iAyN2}
    \end{equation}
which was introduced in \cite{ExpHop2} for neural networks equipped with real-valued neurons, leads for binary neurons $\bm \sigma$ to the same update rule \eqref{eq:s2_1} that we found for the cost function \eqref{eq:hamilt_exp}.  This is due to the monotonicity of the logarithm  in \eqref{eq:H_iAyN2}. Explicitly we can relate the two cost functions by the monotonic relation
\begin{equation}
    \tilde{\mathcal{H}}_N(\bm\sigma|\bm\xi)=- N-\ln\left[-\dfrac{\mathcal{H}_N(\bm\sigma|\bm\xi)}{N}\right].
\end{equation}
\newline   
Therefore, we should expect to obtain similar results when examining \eqref{eq:hamilt_exp} or \eqref{eq:H_iAyN2}, as is indeed corroborated by the numerical evidence provided in Fig.~\ref{fig:s2n_MC_load}.
\end{remark}

\subsection{Patterns as attractors}\label{multiplicative noise}

Having shown that patterns are fixed points of the zero temperature neural dynamics of the network for appropriate $P$, we next investigate the size of their basins of attraction, again as a function of $P$.
To do so, we start the neural dynamics from a corrupted version of the pattern $\bm{\xi}^1$, rather than from the pattern itself as in the previous section. We denote this corrupted version by $\tilde{\bm{\xi}}^1$ and set $\bm{\sigma}^{(0)}= \tilde{\bm{\xi}}^1$, with Mattis overlap
\begin{equation}
    m_1(\tilde{\bm\xi}^1)=\dfrac{1}{N}\SOMMA{i=1}{N}\xi_i^1\tilde{\xi}^1_i= r \in (0, 1]. 
\end{equation}
Equivalently, the Hamming distance (per neuron) between the pattern $\boldsymbol{\xi^1}$ and the input $\boldsymbol{\tilde \xi^1}$  is
\begin{align}
    d \equiv d(\xi^1, \tilde \xi^1)=\dfrac{1}{2}(1-r).
\end{align}
For $r=1$ there is no corruption in the input, while for $r \to 0$ input and pattern become uncorrelated.
\newline
After one step of parallel zero temperature neural updates, i.e. 
\begin{equation}
\begin{array}{lll}
\sigma_i^{(1)}&=&\tilde\xi_i^1\sn\left[ e^{-N}\SOMMA{\mu=1}{P} e^{\sum\limits_{j\neq i}\xi_j^\mu\tilde\xi_j^1}\sinh(\xi_i^\mu)\tilde\xi_i^1\right].
\end{array}
\end{equation}
the value of Mattis magnetization is then
\begin{equation}
\begin{array}{lll}
     m_1^{(1)}&=&\dfrac{1}{N}\SOMMA{i=1}{N}\xi_i^1\tilde\xi_i^1\sn\left[ e^{-N}\SOMMA{\mu=1}{P} e^{\sum\limits_{j\neq i}\xi_j^\mu\tilde\xi_j^1}\sinh(\xi_i^\mu)\tilde\xi_i^1\right]
     \\\\
     &=&\dfrac{1}{N}\SOMMA{i=1}{N}\sn\left[ e^{-N}\SOMMA{\mu=1}{P} e^{\sum\limits_{j\neq i}\xi_j^\mu\tilde\xi_j^1}\sinh(\xi_i^\mu)\xi_i^1\right].
     \label{eq:m1BA}
\end{array}
\end{equation}
Using the Central Limit Theorem again on the sums over $\mu$ and $j$, we can replace each term appearing in the argument of the sign function in \eqref{eq:m1BA} by the Gaussian variable $\mu_1+z_i\sqrt{\mu_2{-\mu_1^2}}$ with $z_i \sim \mathcal{N}(0,1)$, where the moments now involve expectations over both the original patterns and the corrupted initial condition:
\begin{eqnarray}
\mu_1&\coloneqq& \mathbb{E}_{\xi}\mathbb{E}_{\tilde\xi}\left[h_i(\bm\xi^1|\tilde{\bm\xi}^1)\right],\label{eq:mu1}
\\
\mu_2&\coloneqq&\mathbb{E}_{\xi}\mathbb{E}_{\tilde\xi}\left\{\left[ h_i(\bm\xi^1|\tilde{\bm\xi}^1)\right]^2\right\}.\label{eq:mu2}
\end{eqnarray}
Thus, \eqref{eq:m1BA} becomes
\begin{equation} \label{eq:bibo}
    m_1^{(1)}=\dfrac{1}{N}\SOMMA{i=1}{N}\mathrm{sign}\left(\mu_1 +z_i\sqrt{\mu_2-\mu_1^2}\right)\,.
\end{equation}
For large $N$ the arithmetic mean appearing on the right can again be replaced with the expected value, as in \eqref{eq:largemean_matt}.
%that is, given $g(z)=\mathrm{sign}(\mu_1 + z \sqrt{\mu_2{-\mu_1^2}})$, 
% \begin{equation}
%     \dfrac{1}{N}\SOMMA{i=1}{N} g(z_i) \;\;\mathrm{with}\;\; z_i\sim \mathcal{N}(0,1) \xrightarrow[N\to\infty]{} \mathbb{E}[g(z)]=\displaystyle\int \dfrac{dz}{\sqrt{2\pi}}e^{-\frac{z^2}{2}} g(z)\,.
%     \label{eq:largemean}
% \end{equation}
We can therefore rewrite \eqref{eq:bibo} in the large $N$ limit as
\begin{equation}
    m_1^{(1)}=\displaystyle\int\dfrac{dz\,e^{-\frac{z^2}{2}}}{\sqrt{2\pi}} \mathrm{sign}\left(\mu_1+z\sqrt{\mu_2-\mu_1^2}\right)= \mathrm{erf}\left(\dfrac{\mu_1}{\sqrt{2(\mu_2-\mu_1^2)}}\right) \,.\label{eq:m1}
\end{equation}

For the explicit values of \eqref{eq:mu1} and \eqref{eq:mu2} we find
\begin{equation}
\begin{array}{lll}
     \mu_1 =&      e^{-1}\sinh(1)\left[\dfrac{(1+r)+(1-r)e^{-2}}{2}\right]^{N-1},
     \\\\
     \mu_2 =&      e^{-2}\sinh^2(1)\left[\dfrac{(1+r)+(1-r)e^{-2}}{2}\right]^{2(N-1)} +  e^{-2}\sinh^2(1)P \left(\dfrac{1+e^{-4}}{2}\right)^{N-1}.
\end{array}
\end{equation}
Inserting these into \eqref{eq:m1} gives
\begin{equation}
\label{eq:erf}
    m_1^{(1)}=\mathrm{erf}\left\{\left[2P \left(\dfrac{2(1+e^{-4})}{[(1+r)+(1-r)e^{-2}]^2}\right)^{N-1}
  \right]^{-1/2}\right\}.
\end{equation}
Equivalently, if we set 
\begin{align}
    P=\gamma \left(\dfrac{[(1+r)+(1-r)e^{-2}]^2}{2(1+e^{-4})}\right)^{N-1},
    \label{eq:carico_r}
\end{align}
then we can write \eqref{eq:erf} as
\begin{equation}
\begin{array}{lll}
    m_1^{(1)}=\mathrm{erf}\left(\dfrac{1}{ \sqrt{2\gamma }}\right).
\end{array}
\end{equation}
We thus still find an exponential storage capacity guaranteeing pattern recall from corrupted initial conditions, as long as
\begin{equation}
    r> \dfrac{\sqrt{2 (1 + e^4)}}{e^2-1}-\dfrac{1 + e^2}{e^2-1}\circaa 0.337438,
    \label{eq:r}
\end{equation}
or, equivalently, 
\begin{align}
    d<\dfrac{1}{2}-\dfrac{1}{2}\left(\dfrac{\sqrt{2 (1 + e^4)}}{e^2-1}-\dfrac{1 + e^2}{e^2-1} \right)\approx 0.331281.
\end{align}

\begin{figure}
    \centering
    \includegraphics[width=\textwidth]{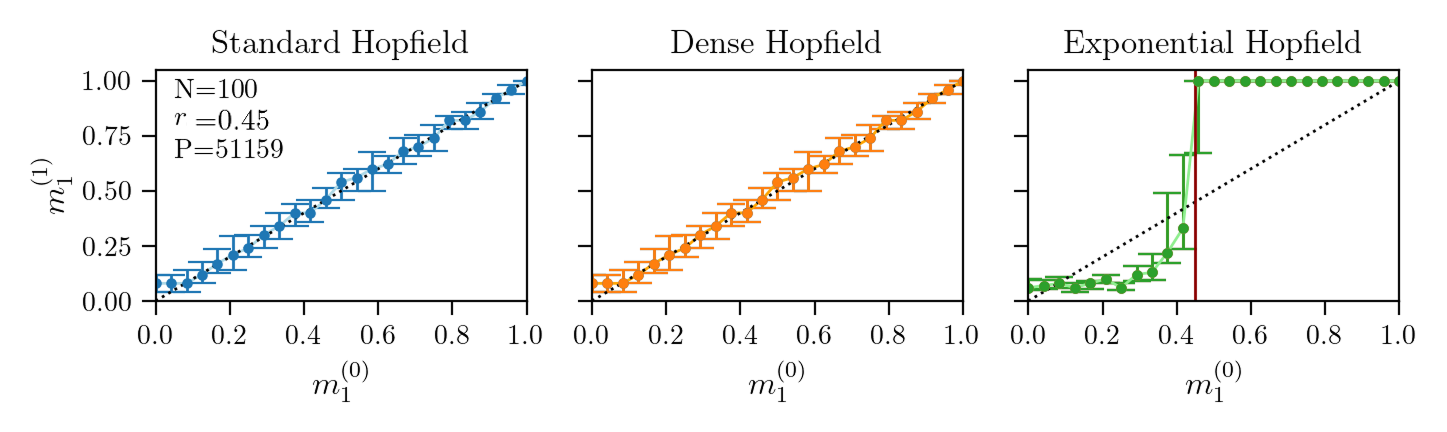}
    \caption{Markov chain Monte Carlo (MCMC) simulations with fixed \( N \) and \( r \) as indicated in the legend. The number of patterns is set according to \eqref{eq:carico_r}. We compare three cases: the standard Hopfield model with pairwise interactions (left panel), the dense Hopfield model with four-body interactions (central panel) and the exponential Hopfield model we study in this paper (right panel).  
\\
    The plots show the Mattis magnetization $m_{\mu}$ pertaining to the retrieved pattern ($\mu=1$) after one parallel MCMC update step, \( m^{(1)}_1 \), as a function of its initial value \( m^{(0)}_1 \). In the right panel, the vertical red line marks the minimum initial magnetization \( m_1^{(0)}=r\) required for perfect retrieval after one MC update as theoretically predicted, showing good agreement with the MCMC results. We note further that, as expected, only the exponential Hopfield network is able to cope with such a huge number of patterns,  while the pairwise and four-body versions achieve no effective stabilization of the pattern ($m_1^{(1)}\approx m_1^{(0)}$).}
    \label{fig:MC_sim}
\end{figure}

Note that if \( r = 1 \), we recover the previous maximal storage defined in \eqref{eq:carico1}. Conversely, if the network needs to handle corrupted patterns as input, its maximal storage capacity must decrease: the greater the noise in the input, the lower the maximal storage capacity. We emphasize that the storage capacity nonetheless remains exponential as shown by \eqref{eq:carico_r}, provided $r$ is large enough (see Eq.~\eqref{eq:r}).
%though still scaling exponentially and respecting the bound on corruption provided in 
%\red{@@ I don't see the point of introducing the new $\otimes$ symbols just to distinguish whether $P$ or $r$ is treated as fixed. I've written this accordingly but left the old version in the latex, commented out. If you agree with my version, the plot legend in Fig 2 would also need to be updated. @@}
Explicitly, perfect recall after one parallel neural update is obtained when $\gamma$ defined in \eqref{eq:carico_r} tends to zero for large $N$. This can be achieved by choosing the number of patterns as e.g.\ 
\begin{equation}
P = \dfrac{1}{\ln(N)} \left(\dfrac{[(1+r)+(1-r)e^{-2}]^2}{2(1+e^{-4})}\right)^{N-1}.
\label{eq:load_r}
\end{equation}
Put differently, for this number of patterns, initial network states with a Mattis magnetization greater than $r$ (or equivalently, a Hamming distance from \( \bm\xi^1 \) smaller than \( d=(1-r)/2 \)) will, after just one step of parallel MCMC updating, lead to a network configuration perfectly aligned with the stored pattern, i.e.\ \( m_1^{(1)} \to 1 \). We demonstrate this numerically in Fig.~\ref{fig:MC_sim}, where we also
%note that in Fig.~\ref{fig:MC_sim} we 
compare to the behavior of the standard Hopfield model (with pair-wise interactions) and a dense Hopfield model with four-body interactions ($p=4$).

\subsection{The {\em perfect recall} regime}\label{NewProof}

In order to corroborate our findings so far regarding the exponential storage capacity of the Hopfield model studied here, we provide a further independent argument based on considering its equilibrium statistics.

\newcommand{\st}{\tilde{\sigma}}
Let us start from the parallel neural dynamics update rule \eqref{eq:s2_1},  which for large $n$ will generate samples from a zero temperature ($\beta\to \infty$) Boltzmann distribution, or from one of its ergodic sectors. Separating off the contribution from pattern $\mu=1$, the update rule reads 
\begin{equation}
    \sigma_i^{(n+1)}
    = \sigma_i^{(n)}\:\sn\left[ e^{-\xi_i^1\sigma_i^{(n)}+N( m_1^{(n)}-1)}\sinh(\xi_i^1)\sigma_i^{(n)}+e^{-N}\SOMMA{\mu>1}{P} e^{\sum\limits_{j\neq i}
    %\xi_j^1\xi_j^\mu\st_j
    \xi_j^\mu\sigma^{(n)}_j}
\sinh(\xi_i^\mu)\sigma_i^{(n)}\right].
    \label{mf2}
\end{equation}
In equilibrium the statistics of $\sigma_i^{(n+1)}$ are identical to those of $\sigma_i^{(n)}$ by construction, so dropping the iteration superscripts we have the equality between equilibrium averages
\begin{equation}
    \langle \sigma_i\rangle
    = \left\langle
    \sigma_i\:\sn\left[ e^{-\xi_i^1\sigma_i+N( m_1-1)}\sinh(\xi_i^1)\sigma_i+e^{-N}\SOMMA{\mu>1}{P} e^{\sum\limits_{j\neq i}
    %\xi_j^1\xi_j^\mu\st_j
    \xi_j^\mu\sigma_j}
\sinh(\xi_i^\mu)\sigma_i\right] \right\rangle\ .
\label{mf2b}
\end{equation}
Setting  $\st_i=\xi_i^1\sigma_i$, Eq.~\eqref{mf2b} can, after multiplication by $\xi_i^1$, be written in the compact form
\begin{equation}
    \langle\st_i\rangle
    = \langle \sn(c\, e^{-\st_i}+z_i)
    \rangle 
    \label{mf_1}
\end{equation}
where 
\begin{align}
    c=&e^{N(m_1-1)}, \label{c_def}\\
    z_i=& e^{-N}\sum_{\mu>1}^P \dfrac{\sinh(\xi_i^1\xi_i^\mu) }{\sinh(1)}\prod\limits_{j\neq i}\exp \left[\xi_j^1\xi_j^\mu\st_j\right],
    \label{z_def}
\end{align}
with $z_i$ playing the role of a noise. Note that as we can explicitly write $c\,e^{-\st_i} = e^{-N+\sum_{j\neq i}\st_j}$, the quantity being averaged on the r.h.s.\ of \eqref{mf_1} is a function only of the $\st_j$ with $j\neq i$.
%with $z_i$ playing as the noise. Now, as we can explicitly write $c\,e^{-\st_i} = e^{-N+\sum_{j\neq i}\st_j}$, the quantity being averaged on the r.h.s.\ of \eqref{mf_1} is a function only of the $\st_j$ with $j\neq i$. 
% Therefore, if we write 
% \begin{equation}
%     \langle \st_i \rangle =\langle \mathrm{sign}(c\,e^{-\st_i} + z_i)\rangle,
%     \label{MF_eq}
% \end{equation}
% we can suppose that the average on the r.h.s. is over all spins $\st_j$ other than $\st_i$. 
\newline
The identity \eqref{mf_1} is exact with the $z_i$ as defined, i.e.\ being functions of the $\st_j$. As an approximation one could treat the $z_i$ as quenched variables, as in the signal-to-noise approach in Sec.~\ref{Patterns as fixed points}. However, already the systematic interaction term in Eq.~\eqref{mf_1} is not of simple mean-field type in the sense that one could treat $m_1$ as a fixed order parameter. This is because, even if in \eqref{mf_1} just one of the $\st_j$ changes sign -- corresponding to a single spin flip during Monte Carlo dynamics -- this will change the systematic interaction term $ce^{-\st_i}$ by a substantial amount, specifically by a factor $e^2$ or $e^{-2}$. Provided that $z_i<0$, the quantity being averaged on the r.h.s.\ of \eqref{mf_1} can therefore easily change sign.
This implies that correlations between spins will not generically be weak even for large $N$, so that mean-field approximations do not apply.

\newcommand{\prob}{\mathbb{P}}

Nonetheless, the regime of perfect recall ($m_1 \to 1$) can be demarcated relatively easily as it corresponds to $c=1$ and $\st_i=1$ for all $i$. This is indeed a solution of~(\ref{mf_1}) provided that $z_i>-e^{-1}$ for all $i$. Taking the $z_i$ to be independent and identically distributed variables and requiring the probability of {\em all} $z_i$ being above $-e^{-1}$ to be close to unity gives the condition
\begin{equation}
[1-\prob(z<-e^{-1})]^N=1-\delta
\end{equation}
with some small $\delta$, or equivalently
\begin{equation}
\prob(z<-e^{-1})=\delta/N.
\label{P_constraint}
\end{equation}
Calculating the moments of the $z_i$ from the relevant quenched averages as in Eqs.~\eqref{eq:momenta_exp}), we find that they have zero mean and 
%If as the distribution of the $z_i$ we take a Gaussian with 
variance $s^2=e^{-2}P [(1+e^{-4})/2]^{N-1}=e^{-2}\gamma$ with $\gamma$ as defined in \eqref{eq:carico_r} (for $r=1$). 
The constraint \eqref{P_constraint} thus translates into 
\begin{equation}
\frac{1}{2}{\mathrm{erfc}}\left(\dfrac{1}{\sqrt{2}se}\right)=\dfrac{\delta}{N}.
\end{equation}
Given the smallness of the r.h.s., \ one can use the asymptotic form of the complementary error function to simplify this to
\begin{equation}
\frac{se}{\sqrt{2\pi}}e^{-1/(2s^2 e^2)}=
\sqrt{\frac{\gamma}{2\pi}}e^{-1/(2\gamma)}=
\delta/N,
\end{equation}
with asymptotic (large N) solution
\begin{equation}
\gamma = \frac{1}{2\ln (N/\delta)} \qquad 
\mbox{or} \qquad 
P\equiv P^*(N) = \frac{1}{2\ln (N/\delta)}\left(\frac{2}{1+e^{-4}}\right)^{N-1}.
\label{Pstar}
\end{equation}
This result is essentially identical to the estimate \eqref{eq:carico1} for the storage capacity provided by  the signal-to-noise analysis, the difference being only in logarithmic factors. The choice of $\delta$ is essentially arbitrary but one could choose e.g.\ $\delta=N^{-a}$ with $a>0$ so that the probability that recall will fail vanishes as $N\to 0$. This would give $[2(1+a)\ln N]^{-1}$ as the prefactor in~(\ref{Pstar}), illustrating the weak dependence on the specific choice of $\delta$. We note finally that the independence of the $z_i$ assumed above is a simplification. This can be removed by a more elaborate calculation along the lines shown in the Appendix but again only modifies the prefactor in~(\ref{Pstar}).

\section{Conclusions}\label{Conclusions}
Recently, since the explosion of social media and the digital era, several major corporations have been struggling with massive data storage,  exploring new frontiers for data handling such as  cooling data centers under the ocean \cite{Oceano1,Oceano2}.  In this context, beyond empirical approaches, a theoretical solution lies in crafting neural networks with massive storage capacity. In this research field we may locate the so-called {\em exponential Hopfield models}, namely modern Hebbian networks that -- if equipped with $N$ neurons -- can store an exponential (in $N$) number of patterns $P$. Such models have appeared in rapid succession in the recent literature, using both binary \cite{ExpHop1} and real-valued neurons \cite{ExpHop2,LucibelloMezard}.

In these short notes we have introduced and studied another cost function for an exponential Hopfield model: based on the (extensive) standard loss functions $\mathcal{L}^{\mu}= \frac{1}{2N}||\boldsymbol{\xi}^{\mu} - \boldsymbol{\sigma}||^2$, one per pattern, we defined the overall cost function of the network 
%--within this {\em exponential context}--  $\mathcal{C}:=
as $-N\sum_{\mu}^P\exp\left(-N \mathcal{L}^{\mu}\right)$, using the exponentials of the $L_2$-loss functions  rather than these loss function themselves. As the exponent can be written as $-N\mathcal{L}^{\mu} = N (m_{\mu}-1)$ in terms of the Mattis magnetizations $m_\mu$, the underlying motivation for this choice of cost function is that minimizing it should force one of the $m_\mu$ towards unity:
%achieve its minimization  --in the large $N$ limit\footnote{Note that, in this asymptotic limit, patterns become orthogonal such that if pattern $\mu=1$ has been retrieved, $m_1 =1$ and all the others $m_{\mu \neq 1}=0$.}-- the network would eventually force the Mattis magnetization to be one even for not-zero values of the exponential storage capacity $\gamma$: 
this  defines the {\em perfect recall regime} that we mainly focussed on. 

The main reward of this variation on a theme is that the resulting network can store an exponential number of patterns\footnote{There %exponential storage is the leading term: there is 
are logarithmic corrections to this,
%take into account for the sharp storage,  
as in all the other exponential models \cite{ExpHop1,ExpHop2} and also in the standard reference \cite{Bovier}.}. Even if we insist on enlarging their basins of attraction to facilitate the reconstruction of patterns from noisy inputs, the storage capacity remains exponential, though of course the relevant exponent decreases.
%exponential  (yet large basins require a significant decrease in the maximal storage).
%\newline

In future work it will be interesting to deepen the role of replica symmetry breaking in the network: very preliminary results seem to suggest that this does not play a major role, at least within the retrieval region\footnote{Indeed there are other Hebbian networks that, in the regime of large number of patterns, behave mainly according to replica symmetry. Examples include the Kanter-Sompolinsky network \cite{KanterSompo} and the sleepy Hopfield model -- or dreaming neural network -- \cite{AlbertDreaming}, where dreaming destroys the spin glass region \cite{AlbertD2}.}. A similar phenomenon occurs in conventional spin glasses. Here the solution for the pairwise scenario captured by the Sherrington-Kirkpatrick model \cite{SKmodel} has full replica symmetry breaking \cite{MPV}, making this model in some sense the ``most glassy'' within its family. In contrast, the random energy model \cite{REM1,REM2} introduced by Derrida, which is the limit of infinite interaction order $p$ and can be viewed as the pure glassy counterpart of the exponential Hopfield  network, only has one step of replica symmetry breaking.

\section{Acknowledgments}
A.A.\ is grateful to Salento University for funding via the PhD-AI program.
\newline
L.A.\ and A.B.\ are grateful to the PRIN 2022 grant {\em Statistical Mechanics of Learning Machines: from algorithmic and information theoretical limits to new biologically inspired paradigms} n.~20229T9EAT funded by European Union -- Next Generation EU. 
\newline
L.A.\ acknowledges fundings also by the PRIN 2022 grant {\em “Stochastic Modeling of Compound Events (SLIDE)”} n.~P2022KZJTZ funded by the Italian Ministry of University and Research (MUR) in the framework of European Union -- Next Generation EU.
\newline
A.B.\ acknowledges fundings also by Sapienza University of Rome via the grant {\em Statistical learning theory for generalized Hopfield models}.
\newline
A.A., L.A.\ and A.B.\ are members of the group GNFM of INdAM, which is also acknowledged.

\section{Appendix: Stability against all spin flips
%Inspecting extensive fluctuations in the signal-to-noise
}
According to \eqref{eq:s2_1}, the pattern $\bm\sigma=\bm\xi^1$ is stable with regard to a single spin flip -- at neuron $i$, say -- when
\begin{equation}
     \Delta E=2Nh_i(\bm \xi^1_{\backslash i} \vert \bm \xi)\xi_i^1 > 0 \quad\Longrightarrow \quad e^{-N}\SOMMA{\mu=1}{P} e^{\sum\limits_{j\neq i}\xi_j^\mu\xi^1_j}\sinh(\xi_i^\mu)\xi^1_i>0.
     \label{ineq2}
\end{equation}
Let us define the random variables $X_i$, $i \in (1,..,N)$ as  
\begin{equation}
    X_i:=h_i(\bm\xi^1_{\backslash i} \vert \bm \xi)\xi_i^1=\SOMMA{\mu=1}{P} e^{\sum\limits_{j\neq i}\xi_j^\mu\xi^1_j}\sinh(\xi_i^\mu)\xi^1_i.
\end{equation}

In the main text we approximated the distribution of $X_i$  by the Gaussian  $\mathcal{N}(\mu_1,\sqrt{\mu_2-\mu_1^2}\;)$, with moments given by
\begin{equation}
\begin{array}{lll}
     \mu_1 = \mathbb{E}\left[e^{-N}\SOMMA{\mu=1}{P} e^{\sum\limits_{j\neq i}\xi_j^\mu\xi^1_j}\sinh(\xi_i^\mu)\xi^1_i\right]=  e^{-1}\sinh(1), 
     \\\\
     \mu_2 =\mathbb{E}\left[\left(e^{-N}\SOMMA{\mu=1}{P} e^{\sum\limits_{j\neq i}\xi_j^\mu\xi^1_j}\sinh(\xi_i^\mu)\xi^1_i\right)^2\right]= e^{-2}\sinh^2(1) +  e^{-2}\sinh^2(1) P \left(\dfrac{1+e^{-4}}{2}\right)^{N-1}.
\end{array}
\label{eq:momenta_exp2}
\end{equation}
Thus 
\begin{equation}
     X_i=\mu_1+ z_i\sqrt{\mu_2-\mu_1^2}, \qquad 
     %\;\;\;\;
     z_i\sim\mathcal{N}(0,1)
     \label{eq:internal_xi}
\end{equation}
such that $X_i >0$ implies 
\begin{equation}
    z >-
    \left[
    P \left(\dfrac{1+e^{-4}}{2}\right)^{N-1}
    \right]^{-1/2}
    %\dfrac{1}{\sqrt{P \left(\dfrac{1+e^{-4}}{2}\right)^{N-1}}}.
\end{equation}
and this happens with probability $\pi$:
\begin{equation}
    \pi = \mathbb P\left(z >-\left[
    P \left(\dfrac{1+e^{-4}}{2}\right)^{N-1}
    \right]^{-1/2}
    \right)=1-\dfrac{1}{2} \mathrm{erfc}\left(
    %\left[2P \left(\dfrac{1+e^{-4}}{2}\right)^{N-1}\right]^{-1/2}
    \frac{1}{\sqrt{2\gamma}}\right)
    \label{noflipprob}
\end{equation}
with $\gamma=P](1+e^{-4})/2]^{N-1}$ as before. This probability goes to $1$ as $N\to\infty$ if and only if the exponential storage condition in \eqref{P_cond} is respected. 

Arguably the criterion of stability against flipping a {\em given} spin $\sigma_i$, as used above, is somewhat simplified. One might want to ask instead about stability against {\em all} spin flips, which requires that {\em all} $X_i$ are positive. The probability $\mathbb{P}(X_i\geq 0 \ \forall\, i)$ for this event
%Yet the argument above is rather raw as it only evaluates the stability under the  flip of a single spin and, here, we want to inspect the stability  also under an extensive fluctuation of $O(N)$ spins. 
%\par\medskip
%This calculation is far from trivial but it can be estimated in two steps:  at first we assume that spins are uncorrelated, thus we simply evaluate $\pi^N$, then we improve this estimate by taking into account also their covariance.
%\par\medskip
%If we assume initially that the $X_i$ are independent from each other, then the probability that {\em none} of the spins will flip is simply $\pi^N$. We require $\pi^N\to 1$ for $N\to\infty$, or $\pi=1-o(1/N)$. 
%Let us start by evaluating $\pi^N$: if we assume that the spin flip probabilities are independent, we must evaluate the probability for {\em none} of the spins to flip, namely we require 
%For this scenario to hold, it is enough to have 
%\begin{equation}
%    P=\gamma \frac{1}{\delta_1\log(N)}\left(\frac{2}{1+e^{-4}}\right)^{N-1}\;\;\;\mathrm{with}\;\; \delta_1\geq 2.
%\end{equation} 
%the probability that all the random variables $X_i$ are positive, i.e.\ 
is the average of $\prod_{i=1}^N \Theta(X_i)$, where $\Theta$ is the Heaviside step function.

Evaluating this average evidently requires us to account for correlations between the $X_i$, which within the Gaussian approximation are determined by the covariances. 
Proceeding then as in the calculation of $\mu_2$ in the main text one finds, by exploiting the independence of the pattern entries, for a generic second moment with $i\neq k$
\begin{equation}
\mathbb{E}[X_i X_k]
= e^{-2N}\SOMMA{\mu,\nu=1}{P} 
\mathbb{E}
\left[e^{\sum\limits_{j\neq i,k}\xi_j^\mu\xi^1_j + \xi_j^\nu\xi^1_j}\right]
\mathbb{E}\left[ 
e^{\xi^\nu_i\xi^1_i}
\sinh(\xi_i^\mu)\xi^1_i
\right]
\mathbb{E}\left[ 
e^{\xi^\mu_k\xi^1_k}
\sinh(\xi_k^\nu)\xi^1_k
\right].
\end{equation}
One can check that the product of the last two factors vanishes when $\mu\neq\nu$. Of the remaining cases, $\mu=\nu=1$ gives exactly $\mu_1^2=\mathbb{E}[X_i]\mathbb{E}[X_k]$, while the contribution from $\mu=\nu=2,\ldots,P$ gives the actual covariance as (using again $P-1\approx P$)
\begin{equation}
\mathbb{E}[X_i X_k]
- \mathbb{E}[X_i]\mathbb{E}[X_k] = P \left(\frac{1+e^{-4}}{2}\right)^{N-2}e^{-4}\sinh^4(1) =: c_1.
\end{equation}
%Now we extend our reasoning to include also the covariance $\mathrm{Cov}[X_i, X_k]$. As far as the quantities $X_i$ are correlated,  for $i\neq k$ we have
%\begin{equation}
%c:=\mathrm{Cov}[X_i, X_k]
%= e^{-4}\sinh^{4}(1) P \left(\dfrac{1+e^{-4}}{2}\right)^{N-2}
%\end{equation}
%and 
We can therefore generalize the Gaussian representation \eqref{eq:internal_xi} to one for the {\em joint} distribution of the $X_i$:
\begin{equation}
\label{eq:Xi_rep}
\begin{array}{lll}
         X_i &\sim& \mu_1 + z_0 \sqrt{c_1} + z_i \sqrt{\mu_2-\mu_1^2-c_1}\;\;\;\mathrm{with}\;\;z_i,z_0\sim\mathcal{N}(0,1), \qquad i=1, \hdots, N. 
\end{array}
\end{equation}
The average over the $z_i$ of $\prod_{i=1}^N \Theta(X_i)$ then still factorizes and gives
%Now we need to evaluate the probability that all the random variables $X_i$ are positive, i.e.\ the average of $\prod_{i=1}^N \Theta(X_i)$, where $\Theta$ is the Heaviside-Theta function. Using the representation \eqref{eq:Xi_rep} and integrating out the factorized  $z_i$ variables we get for each $i$ an error function
\begin{equation}
    \mathbb{P}(X_i\geq0\ \forall\,i)=\left[1-\dfrac{1}{2}{\rm erfc}\left(\dfrac{\mu_1+z_0\sqrt{c_1}}{\sqrt{2c_2}}\right)\right]^N\ ,
\end{equation}
where $c_2:=\mu_2-\mu_1^2-c_1$ and the r.h.s.\ still needs to be averaged over $z_0$. In order for the resulting probability to be close to unity, we need the erfc to be small for typical $z_0$ of $O(1)$ so can use its asymptotic form for large $\mu_1/\sqrt{2c_2}$ on the r.h.s.\ to obtain
\begin{equation}
    \mathbb{P}(X_i\geq0\ \forall\,i)=1-N\frac{\sqrt{c_2}}{\mu_1\sqrt{2\pi}}
    e^{-(\mu_1+z_0\sqrt{c_1})^2/(2c_2)}\ .
\end{equation}
The average over $z_0$ can now be performed and gives
\begin{equation}
    \mathbb{P}(X_i\geq0\ \forall\,i)=1 - 
\dfrac{N c_2 e^{-\frac{\mu_1^2}{2c_2}}}{\mu_1\sqrt{2\pi(c_2+c_1)}} 
= 1 - \dfrac{N (\mu_2-\mu_1^2-c_1) e^{-\frac{\mu_1^2}{2c_2}}}{\mu_1\sqrt{2\pi(\mu_2-\mu_1^2)}} .
\label{P_all_Xi_positive}
\end{equation}
Note that ignoring correlations in this calculation would correspond to setting $c_1=0$. In that case the average of $\prod_{i=1}^N \Theta(X_i)$ would factorize and therefore give $\pi^N$, with $\pi$ as calculated above. Given that $c_1$ enters with an overall positive sign in (\ref{P_all_Xi_positive}), we thus have the intuitively reasonable bounds
\begin{equation}
    \pi^N < \mathbb{P}(X_i\geq0\ \forall\,i) \leq  \pi
\end{equation}
where the second one follows trivially from the fact that $\prod_{j=1}^N \Theta(X_j)\leq \Theta(X_i)$ for any fixed $i$. Without going further, one thus already sees that the effects of requiring stability to all spin flips rather than just to a single given flip can essentially be understood by replacing $\pi$ by $\pi^N$, or equivalently by replacing the criterion $\pi\to 1$ for pattern stability by the somewhat stronger $\pi=1-o(1/N)$.

More quantitatively, one finds by inserting the values of $\mu_1$, $\mu_2$ and $c_1$ that
\begin{equation}
    \mathbb{P}(X_i\geq0\ \forall\,i)=1 - 
N c_3 \sqrt{\frac{\gamma}{2\pi}}e^{-1/(2\gamma)}
, \qquad 
c_3 = 1-\frac{\sinh^2(1)}{\cosh(2)}
\label{P_all_Xi_positive_explicit}
\end{equation}
with $\gamma=P[(1+e^{-4})/2]^{N-1}$ as before. Ignoring correlations, i.e.\ approximating the probability as $\pi^N$ gives $c_3=1$ instead; this result is consistent with the asymptotic expansion of $\pi^N$ for small $\gamma$, using~(\ref{noflipprob}). Either way, one deduces that stability against all spin flips requires $N\gamma^{1/2}e^{-1/(2\gamma)}\to 0$ for $N\to\infty$. Setting this quantity equal to some small $\delta$ gives $\gamma=1/[2\ln(N/\delta)]$ to leading order. The specific choice $\delta=N^{-a}$ yields $\gamma=1/[2(1+a)\ln(N)]$ or equivalently
\iffalse
As we need the product of all the $X_i$, taking to the power $N$, averaging over $Z_0$ and expanding for large argument yields 
\begin{equation}
\begin{array}{lll}
     &&\mathbb{E}_{Z_0}\prod_{i=1}^N\mathbb{P}\left( X_i>0\right)= 1-\dfrac{N c_2 e^{-\frac{\mu_1^2}{2c_2}}}{\mu_1\sqrt{2\pi(c_2+c)}} 
     \\\\
     &&= 1-N\sqrt{\dfrac{1}{2\pi}P\left(\frac{1+e^{N-1}}{2}\right)^{N-1}}\left(1-\frac{\sinh^2(1)}{\cosh(2)}\right)\exp\left[- \left(2P\left(\frac{1+e^{N-1}}{2}\right)^{N-1}\left(1-\frac{\sinh^2(1)}{\cosh(2)}\right)\right)^{-1}\right].
\end{array}
\end{equation}
So the condition for stability is that the second term of the right hand side of \eqref{pin_explicit} vanishes as $\mathcal{O}(N^{-\delta_2})$ for $N\to\infty$ with $\delta_2>0$. 
This request is satisfied as long as 
\fi
\begin{equation}
    P = \dfrac{1}{2 (1+a)\ln(N)} \left(\dfrac{2}{1+e^{-4}}\right)^{N-1}.
    \label{Pbound_improved}
\end{equation}
We observe that this is identical to (\ref{Pstar}), demonstrating that the stability analysis and the perfect recall approach at equilibrium are fully consistent with each other.

We note finally that using $\pi$ rather than $\pi^N$ (or indeed the full~(\ref{P_all_Xi_positive_explicit})), i.e.\ considering stability against flips of a fixed spin only, amounts for large $N$ only to dropping the factor of $N$ in~(\ref{P_all_Xi_positive_explicit}). This only has the effect of replacing $1+a$ in (\ref{Pbound_improved}) by $a$, thus recovering (\ref{eq:carico1}) in the main text as expected. Overall, we conclude that the main finding of an exponential storage capacity is, up to prefactors, insensitive to the specific choice of stability criterion.
%which we had derived by looking only at stability against single spin flips. Looking back, one sees that stability against all spin flips only reduces the prefactor in $P$, from $1/[2a\ln(N)]$ (which is 
%whatever the resolution level adopted, the storage of patterns in this network is kept exponential in the volume of the neurons $N$.

\end{document}